\documentclass[sigconf, nonacm]{acmart}

\AtBeginDocument{%
  }

\usepackage{subcaption}
\usepackage{tikz}
\usepackage{listings}
\usepackage{xcolor}
\newcommand{\systemnamenf}{Jovis}
\newcommand{\systemname}{{\em Jovis }}
\newcommand{\systemnamens}{{\em Jovis}}

\newcommand{\ie}{{\em i.e.}}
\newcommand{\eg}{{\em e.g.}}

\def\code#1{\texttt{#1}}

\newcommand{\keyword}[1]{{\textit{\textbf{#1.}}}}

\makeatletter
\newcommand*{\rom}[1]{\romannumeral #1}
\newcommand*{\Rom}[1]{\expandafter\@slowromancap\romannumeral #1@}
\makeatother

\theoremstyle{definition}
\newtheorem{example}{Example}

\setcopyright{acmlicensed}
\copyrightyear{2018}
\acmYear{2018}
\acmDOI{XXXXXXX.XXXXXXX}
\acmConference[Conference acronym 'XX]{Make sure to enter the correct
  conference title from your rights confirmation email}{June 03--05,
  2018}{Woodstock, NY}
\acmISBN{978-1-4503-XXXX-X/2018/06}

\begin{document}
\title{\systemnamenf: A Visualization Tool for PostgreSQL Query Optimizer}

\def\snudept{Computer Sci \& Eng}
\def\snu{Seoul National University}
\def\snucity{Seoul}
\def\snucountry{Korea}
\def\snuaffl{
  \institution{\snu} \city{\snucity} \country{\snucountry}
}

\author{Yoojin Choi}
\affiliation{\snuaffl}
\email{cyj@dbs.snu.ac.kr}

\author{Juhee Han}
\affiliation{\snuaffl}
\email{juheehan@dbs.snu.ac.kr}

\author{Kyoseung Koo}
\affiliation{\snuaffl}
\email{koo@dbs.snu.ac.kr}

\author{Bongki Moon}
\affiliation{\snuaffl}
\email{bkmoon@snu.ac.kr}

\begin{abstract}

Query optimizers are essential components of relational database management systems that directly impact query performance as they transform input queries into efficient execution plans.
While users can obtain the final execution plan using the \code{EXPLAIN} command and leverage existing visualization tools for intuitive understanding, the internal decision-making processes of query optimizers are hidden from users, making it difficult to understand how the plan is constructed.
To address this challenge, we present \systemnamens, an interactive visualization tool designed to explore the query optimization process in PostgreSQL.
\systemname provides a comprehensive view of the entire optimization workflow through tailored visualization for each optimization strategy.
It also includes features that allow users to participate in optimization by providing hints, tuning parameters, and reusing prior optimization results.
\systemname serves as both an educational tool for learners and a practical resource for database professionals, helping users understand and improve query optimization by guiding the optimizer to make better decisions or consider previously unexplored plans.
The source code, data, and/or other artifacts have been made available at \url{https://github.com/orgs/snu-jovis}.

\end{abstract}

\ccsdesc[500]{Information systems~Data management systems}
\ccsdesc[500]{Human-centered computing~Visualization systems and tools}

\keywords{Query Optimization, Interactive Visualization, PostgreSQL, Database Management Systems}


\maketitle

\section{Introduction}

Query optimizers are a core component of relational database management systems (RDBMS) as they play a key role in determining query performance.
They transform declarative SQL queries into efficient execution plans by exploring a vast search space and selecting the most optimal plan.
At the same time, query optimization is one of the most challenging aspects of DBMS design, primarily due to the NP-hard nature of finding an optimal join order~\cite{np-hard} and the complexity of picking the best physical operators for each logical operation.
These challenges have driven decades of research, from the traditional cost-based optimizers such as System R~\cite{systemR} to recent machine learning-based approaches~\cite{bao, balsa}.

Understanding the decision-making process of the query optimizer greatly benefits both learners and practitioners.
For learners, it provides a grasp of query optimization beyond the high-level descriptions of execution plans, such as how join orders and physical operators are determined.
For practitioners, deeper insight into how plans are generated, including which alternatives were considered, how the costs were estimated and compared, and the rationale behind the final plan, helps explain suboptimal decisions and enables manual tuning using techniques like hints or configuration adjustments.
Furthermore, such transparency is essential for extending or improving the optimizer as it clarifies its underlying strategies and implementation.

Despite its importance, the query optimization process remains largely opaque to users unless they delve into the source code.
While DBMSs like PostgreSQL~\cite{postgresql} and MySQL~\cite{mysql} provide \code{EXPLAIN} and \code{ANALYZE} commands to display the final execution plan and its estimated cost, along with external tools PEV2~\cite{pev2} and MySQL Workbench~\cite{workbench} that offer graphical representations of the plan, these outputs only reflect the outcome of the optimization process rather than the reasoning behind it.
Several research efforts have aimed to make query optimization more accessible through visualization and interactive features.
QO-Insight~\cite{qo-insight} focuses on visualizing query execution traces of steered query optimizers, which use hints to correct planning mistakes and MOCHA~\cite{mocha} enables users to analyze the impact of alternative physical operator choices.
While these works visualize and help compare the query execution plans, they fall short of providing a comprehensive view of the internal decision-making process of query optimizers.


In this demonstration, we present \systemnamens, an interactive visualization tool designed to reveal the query optimization process of PostgreSQL.
Unlike existing tools that focus solely on the final execution plan, \systemname provides visibility into \textit{how} the optimizer constructs the plan.
It visualizes the explored search space, the evaluated join orders and physical operators, and the cost estimations that drive plan selection.
Beyond visualization, \systemname introduces user-guided optimization, allowing users to steer the optimizer toward better decisions or explore untapped search spaces.
By making the optimizer transparent and interactive, \systemname not only bridges the gap between high-level execution plans and the underlying optimization logic but also helps users diagnose suboptimal choices and improve query performance.

Key features of \systemname include:
\begin{itemize}
    \item Comprehensive visualizations of PostgreSQL's optimization process, covering both its cost-based optimizer using dynamic programming and genetic algorithm.
    \item An in-depth view of the cost model, including cost estimation formulas that illustrate how costs impact plan selection.
    \item User-guided optimization, an interactive feature that allows users to guide the optimizer through hints, configuration adjustments, and leveraging prior optimization results.
\end{itemize}
We chose PostgreSQL as our starting point due to its wide adoption and two complementary optimization approaches.
Our tool is designed to be extensible, with future work planned to support additional DBMSs.


\section{Related Work}

\keyword{Visualization of Query Optimization}
Several systems address specific challenges in visualizing query optimization in the database community.
Picasso~\cite{picasso} provides graphical profiling and analysis of query optimizer behavior.
QE3D~\cite{qe3d} introduces a three-dimensional representation of distributed query plans, helping users identify key performance aspects.
DBinsight~\cite{dbinsight} visualizes the query processing pipeline and unifies heterogeneous data structures across RDBMSs through a common interface.
MOCHA~\cite{mocha} explores the impact of physical operator selection by allowing users to compare alternative plans and observe corresponding performance changes.
QO-Insight~\cite{qo-insight} targets steered query optimizers that use hints to address planning mistakes, offering a visual inspection of query execution traces.

A variety of visualization tools for query optimizers have been developed.
Postgres Explain Visualizer (PEV)~\cite{pev} and its successor PEV2~\cite{pev2} render \code{EXPLAIN} output into interactive trees, enabling users to analyze the structure and estimated costs of execution plans.
Similarly, MySQL Workbench~\cite{workbench} provides graphical plan visualization for MySQL as part of a broader suite of database modeling and administration tools.

Efforts have also been made to visualize genetic algorithm~\cite{cruz}, which inspired \systemnamens's approach to GEQO visualization.
However, our system is tailored specifically to GEQO, leveraging unique data extracted from optimizer logs and offering interactive features for genetic query optimization.
To the best of our knowledge, \systemname is the first tool to provide dedicated visualization for GEQO.

\keyword{PostgreSQL Query Optimizer}
In PostgreSQL, the optimizer examines possible query plans by generating data structures called \textit{paths}, which represent partial or complete plans along with their estimated costs.
The standard optimizer builds these paths by enumerating possible plans using dynamic programming~\cite{dp} and evaluates them with a cost model to determine the cheapest path.
However, this approach becomes infeasible for queries involving a large number of joins, as exploring the exponentially growing search space requires excessive time and memory. 
In such cases, PostgreSQL switches to the Genetic Query Optimizer (GEQO), which applies a genetic algorithm to iteratively generate and refine join sequences over successive generations using recombination mechanisms, balancing optimization quality with computational efficiency~\cite{pg-doc-geqo}.

\section{\systemname}

\begin{figure}[t]
    \centering
    \includegraphics[width=\linewidth]{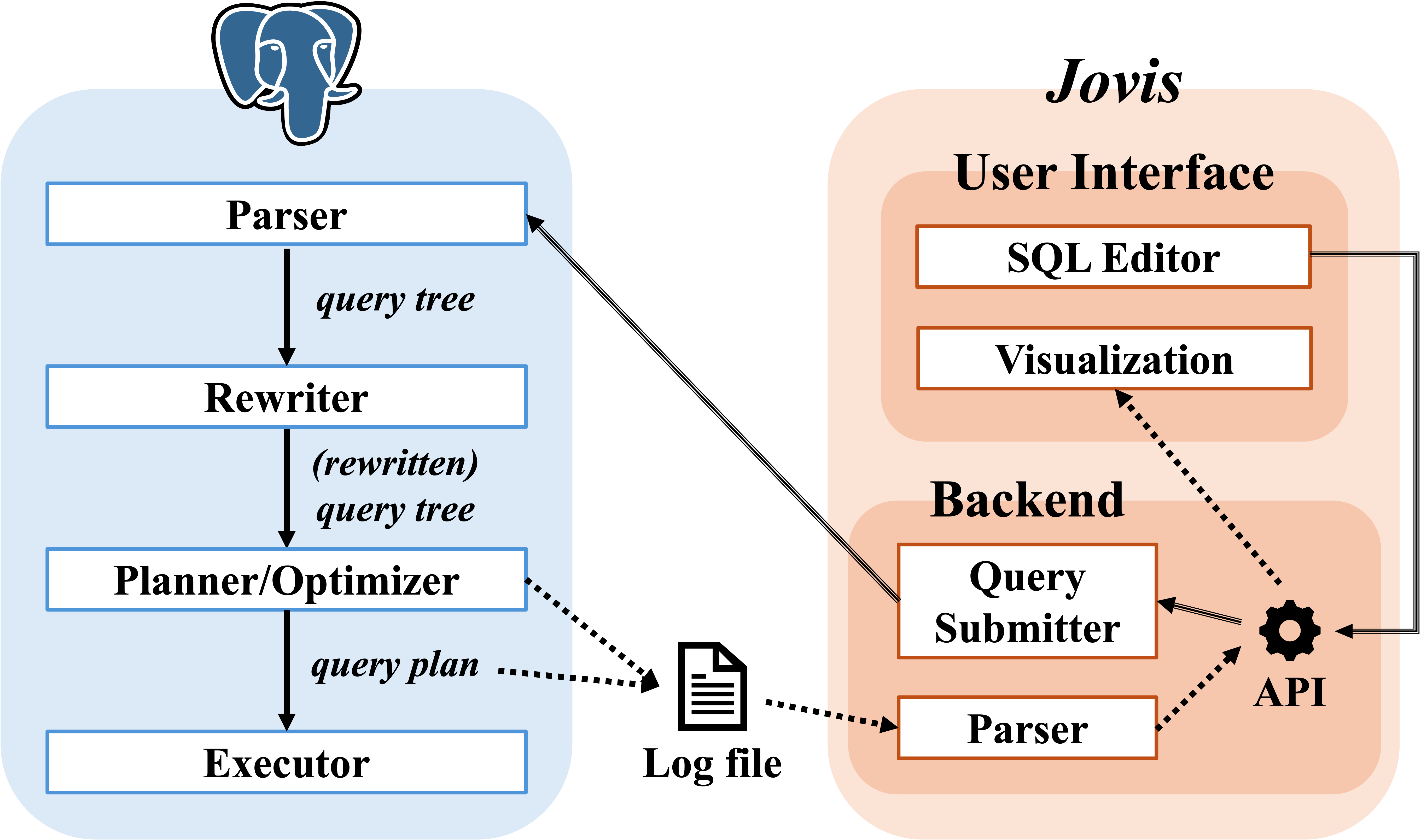}
    \caption{Architecture of \systemname}
    \label{fig:arc}
\end{figure}

\begin{figure*}[t]
    \centering
    \begin{tikzpicture}
        \node[inner sep=0pt, rounded corners=5pt, clip]
            {\includegraphics[width=\linewidth]{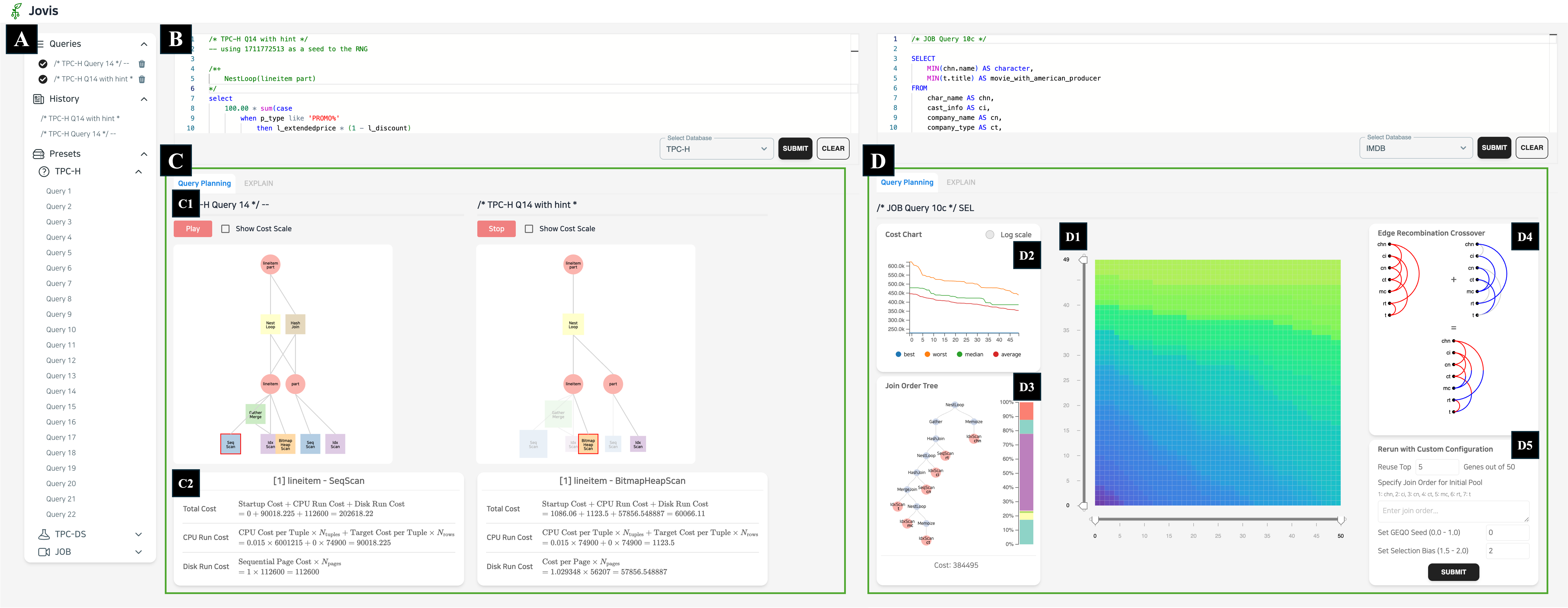}};
    \end{tikzpicture}
    \caption{Graphical User Interface of \systemname}
    \label{fig:full}
\end{figure*}

\begin{figure}[t]
    \centering
    \begin{tikzpicture}
        \node[inner sep=0pt, rounded corners=5pt, clip]
            {\includegraphics[width=0.95\linewidth]{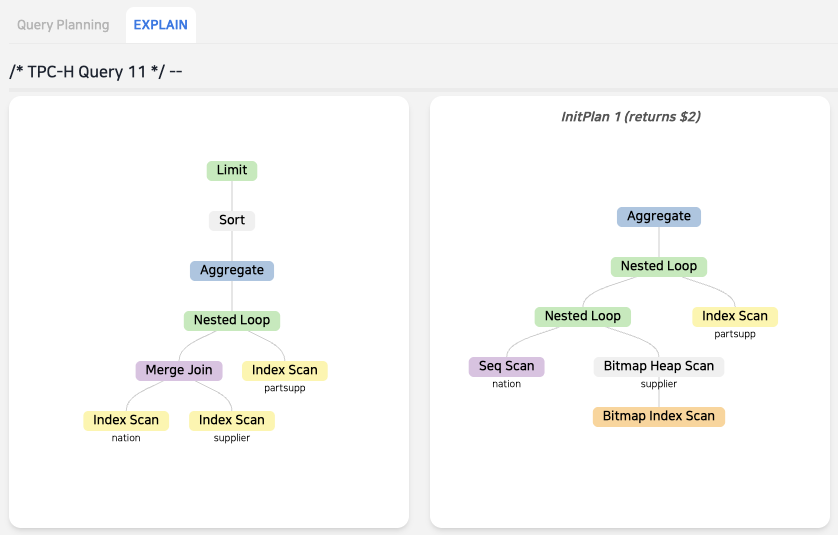}};
    \end{tikzpicture}
    \caption{Visualization of \code{EXPLAIN} for TPC-H Query 11}
    \label{fig:explain}
\end{figure}

This section provides an overview of the architecture of \systemnamens, illustrated in Figure~\ref{fig:arc}.
\systemname consists of three main components: a patched version of PostgreSQL, an interactive user interface, and a backend.
Users submit queries through the frontend's SQL editor, which are transmitted to PostgreSQL via the backend.
PostgreSQL processes the queries and generates detailed optimizer logs, which are collected and parsed by the backend into structured JavaScript Object Notation (JSON).
The processed data is then sent to the frontend, where it is visualized through an interactive interface.
We applied a lightweight patch to PostgreSQL 16.2 to add detailed logging of the optimizer’s internal decisions, including paths and costs.
Since the patch is non-intrusive, the system can be easily extended to other PostgreSQL versions.

\subsection{Graphical User Interface}
Figure~\ref{fig:full} represents the graphical user interface (GUI) designed to offer an intuitive and interactive experience for users to explore and understand the query optimization process.
Built with React and D3.js~\cite{d3}, the interface leverages data-driven visualizations to present complex optimization details in an accessible manner.

\textbf{Sidebar} \colorbox{black}{\textcolor{white}{A}} provides a flexible interface for managing query planning visualizations.
It consists of three components:
(\rom{1}) \textbf{Queries} allows users to select, deselect, or delete submitted queries, supporting side-by-side comparison of multiple queries and their optimizations.
This helps users analyze how different queries or configurations affect the optimizer's decisions, (\rom{2}) \textbf{History} maintains a record of previously executed queries, enabling quick reuse and reducing repetitive input and (\rom{3}) \textbf{Presets} includes predefined queries from widely used benchmarks, TPC-H~\cite{tpch}, TPC-DS~\cite{tpcds}, and Join Order Benchmark (JOB)~\cite{job}.
It simplifies experimentation and performance evaluation.
\textbf{SQL editor} \colorbox{black}{\textcolor{white}{B}} allows users to submit queries, select a database, and execute them for analysis.

For both the query planning views \colorbox{black}{\textcolor{white}{C}} and \colorbox{black}{\textcolor{white}{D}}, users can switch to the \code{EXPLAIN} tab to visualize the \code{EXPLAIN} output as a hierarchical tree as presented in Figure~\ref{fig:explain}.
This view provides an intuitive understanding of the final execution plan.

\begin{figure}[t]
    \centering
    \begin{tikzpicture}
        \node[inner sep=0pt, rounded corners=5pt, clip]
            {\includegraphics[width=0.9\linewidth]{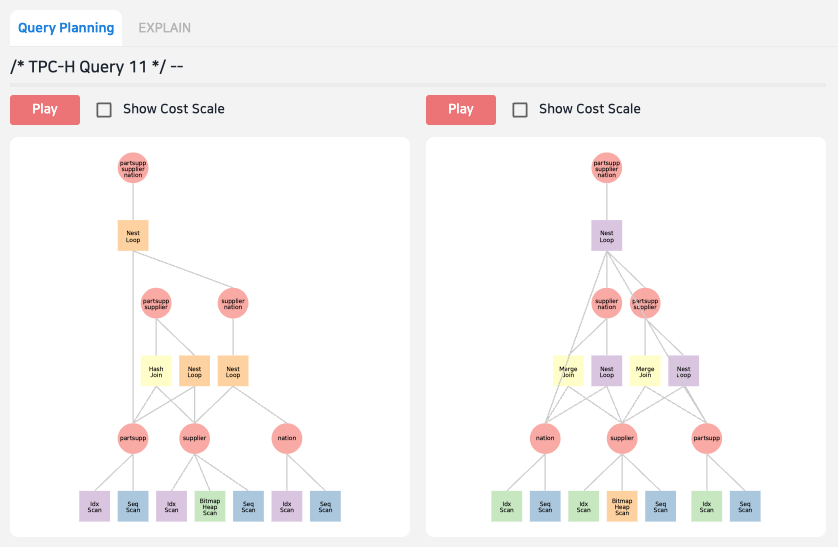}};
    \end{tikzpicture}
    \caption{Subquery Support in TPC-H Query 11}
    \label{fig:subquery}
\end{figure}

\subsubsection{Standard Optimizer}
\textbf{Query Planning View} \colorbox{black}{\textcolor{white}{C}} visualizes PostgreSQL's standard optimization process using dynamic programming.
The optimizer adopts a bottom-up approach, starting by determining access paths for each base relation.
It then incrementally constructs join sequences and selects physical operators until the complete execution plan is achieved.

To effectively depict the process, \systemname employs a directed acyclic graph (DAG), which is well-suited for representing the bottom-up strategy due to its ability to handle multiple parent nodes and hierarchical relationships.
In this structure, circular nodes represent relations or join sequences, while rectangular nodes represent physical operators considered during plan enumeration.
Edges indicate the application of an operator to its input relations or join orders.
The DAG is presented in an inverted layout where operators appear as children of their associated relations.
For example, the left DAG in \colorbox{black}{\textcolor{white}{C1}} shows multiple scan methods evaluated for each relation, and nested loop join and hash join were explored for joining two relations.
Furthermore, as shown in Figure~\ref{fig:subquery} for TPC-H Query 11, \systemname supports subqueries by generating separate DAGs for each one, making it possible to inspect complex queries in a modular way.

Users can interact with the DAG through several features.
The \textbf{Play/Stop button} in \colorbox{black}{\textcolor{white}{C1}} animates the optimization process, reducing the opacity of non-selected nodes to make the decision path easy to track.
The \textbf{Cost Scale toggle} adjusts the size of operator nodes based on their estimated cost, highlighting why specific operators or join sequences were chosen.
The right DAG shows the result after activating both features.
Additionally, \textbf{clicking on an operator node} reveals detailed cost factors and calculation formulas in \colorbox{black}{\textcolor{white}{C2}}.

\begin{example}
    \colorbox{black}{\textcolor{white}{C2}} illustrates an example of cost analysis comparing a sequential scan and a bitmap heap scan for the \code{lineitem} relation in TPC-H Query 14.
    The query needs to access 74,900 tuples out of a total of 6,001,215 ($N_{rows}$).
    Although the bitmap heap scan incurs a startup cost for building the list of disk pages to fetch, it significantly reduces the number of pages ($N_{pages}$) and tuples ($N_{tuples}$) accessed.
    This reduction in disk I/O and CPU costs for scanning only the relevant records makes the bitmap heap scan more efficient than a sequential scan, which fetches all pages and processes every record.
    Due to the query's low selectivity, the bitmap heap scan achieves a much lower cost, leading the optimizer to select it.
    As demonstrated in this example, users can understand the rationale behind the optimizer's decision by examining the cost.
\end{example}

By combining these features, \systemname helps users visually grasp all paths considered during optimization and the rationale behind the optimizer's decisions, including detailed cost calculations.

\begin{figure}[t]
    \centering
     
    \begin{subfigure}{0.5\columnwidth}
        \begin{tikzpicture}
            \node[inner sep=0pt, rounded corners=5pt, clip]
                {\includegraphics[width=\linewidth]{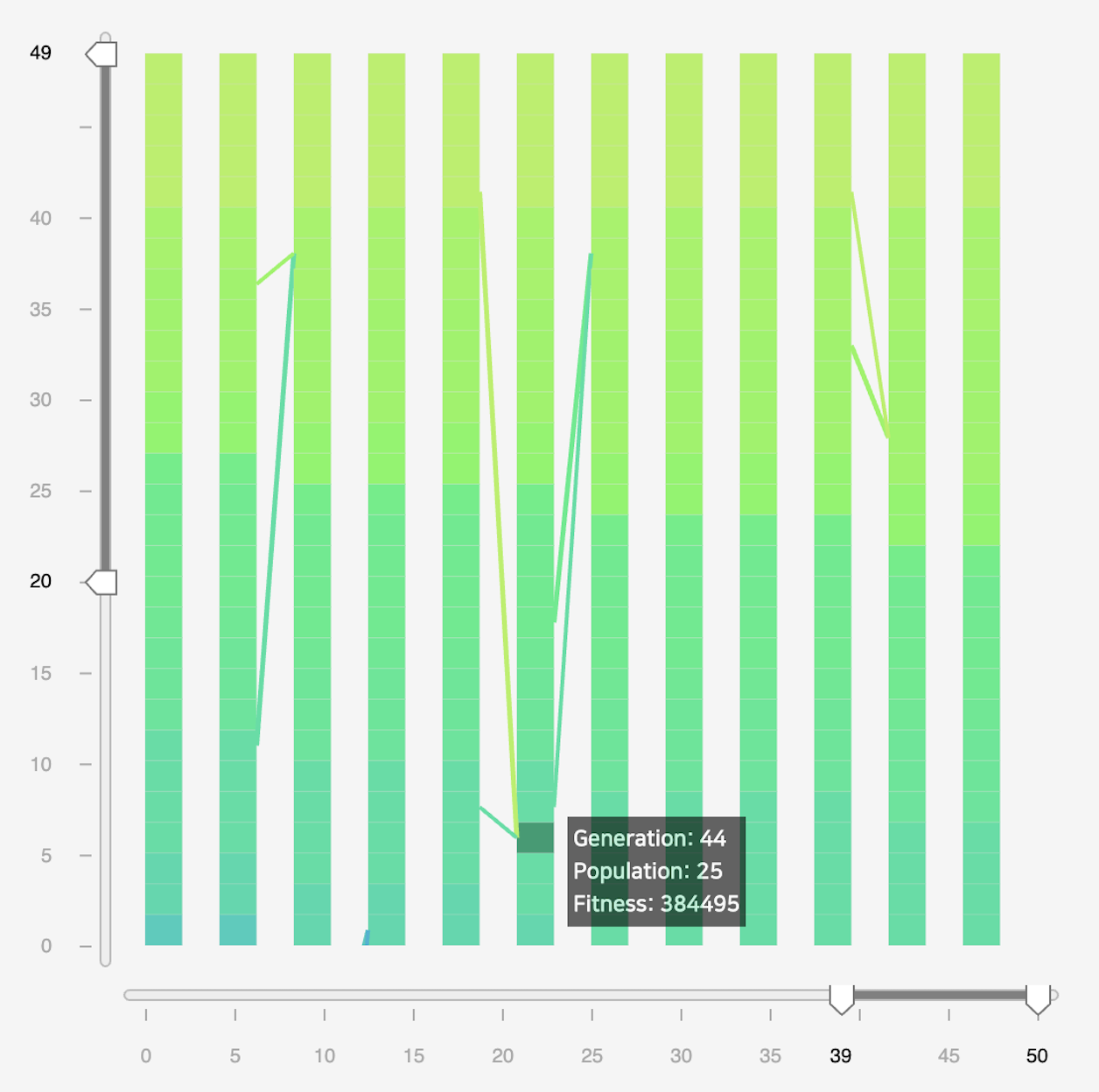}};
        \end{tikzpicture}
        \caption{Range Sliders}
        \label{fig:geqo-range}
    \end{subfigure}
    \hspace{0.5em}
    \begin{subfigure}{0.4\columnwidth}
        \begin{tikzpicture}
            \node[inner sep=0pt, rounded corners=5pt, clip]
            {\includegraphics[width=\linewidth]{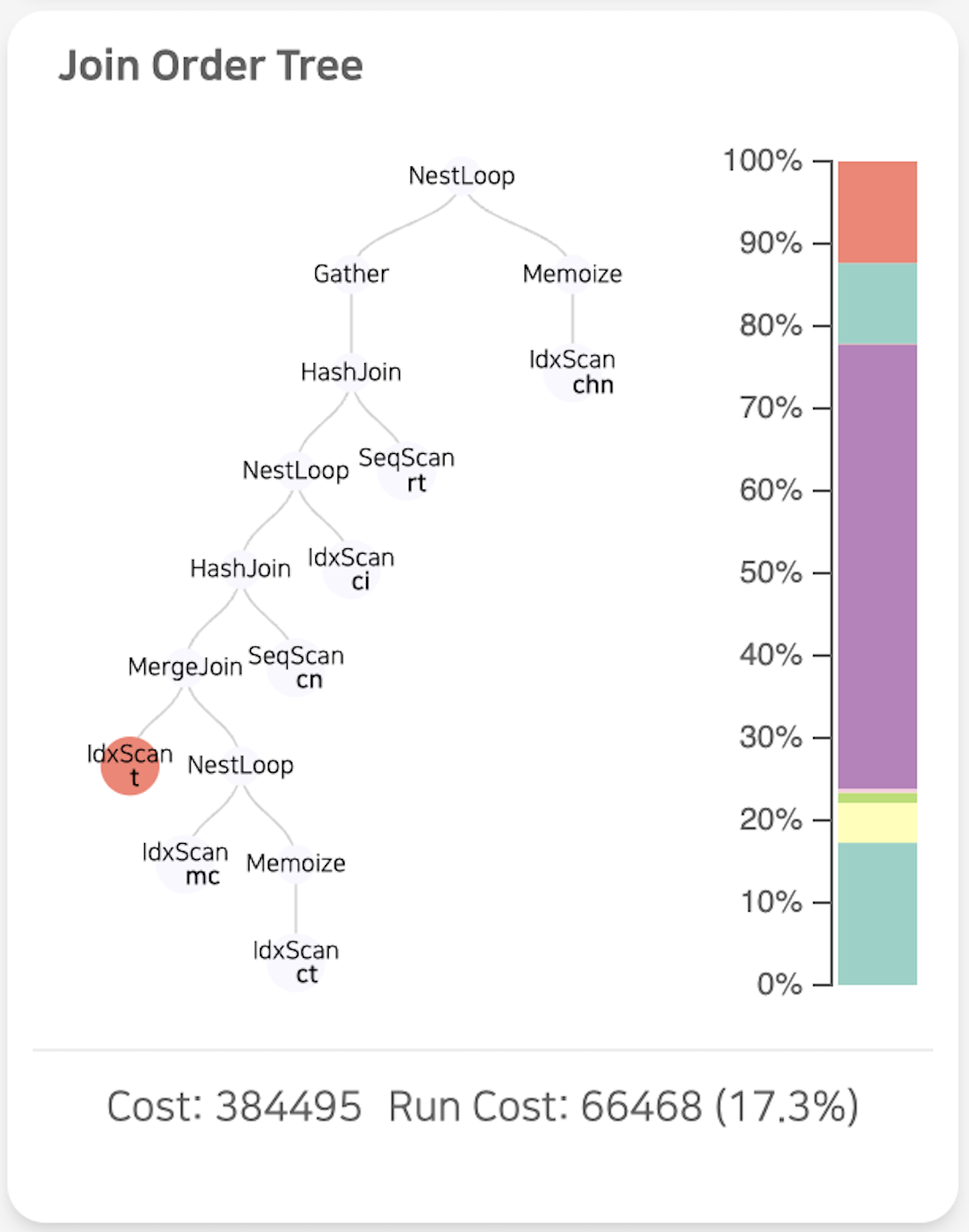}};
        \end{tikzpicture}
        \caption{Stacked Bar Chart}
        \label{fig:geqo-stack}
    \end{subfigure}
    
    \caption{Interactive Features for GEQO}
    \label{fig:demo-geqo}
\end{figure}

\subsubsection{Genetic Query Optimizer}
\textbf{Query Planning View} \colorbox{black}{\textcolor{white}{D}} visualizes PostgreSQL's GEQO, which applies a genetic algorithm to heuristically explore join orders when exhaustive enumeration becomes impractical (\ie, when the number of relations involved in joins reaches or exceeds the PostgreSQL's configurable parameter \code{geqo\_threshold}).
GEQO iteratively refines a population of candidate join sequences, referred to as \textit{genes}, through selection, crossover, and replacement.
Each generation maintains a pool of genes, where lower-cost genes are considered more fit.
New genes are generated via crossover, combining segments of two parent genes to produce offspring.
The pool is then sorted by estimated cost and the least fit genes are discarded.
This process continues until a preset number of join sequences are evaluated.

\systemname employs a grid heatmap, inspired by Cruz~\cite{cruz}, to visualize the GEQO process.
The \textbf{heatmap} \colorbox{black}{\textcolor{white}{D1}} represents the pool of genes across generations, with color gradients to indicate their estimated costs.
This effectively demonstrates how GEQO improves fitness over time.
In addition, a \textbf{line chart} \colorbox{black}{\textcolor{white}{D2}} tracks the best, worst, median, and average costs throughout the optimization process, providing an overview of cost trends.

Users can interact with the visualization to explore the optimization process in detail.
The \textbf{range sliders} in \colorbox{black}{\textcolor{white}{D1}} allow users to focus on specific generations or genes.
As shown in Figure~\ref{fig:geqo-range}, when zoomed in on a smaller range, crossover visualizations appear, illustrating which parent genes contribute to offspring through edges.
In the \textbf{cost chart} \colorbox{black}{\textcolor{white}{D2}}, users can hover over lines to pinpoint the specific generation, toggle between linear and logarithmic scales, and click on legends to isolate metrics for focused analysis.
\textbf{Clicking on a gene} in \colorbox{black}{\textcolor{white}{D1}} displays its join sequence as a tree in \colorbox{black}{\textcolor{white}{D3}}, with operator costs represented as a stacked bar chart.
Users can click on each stack to see how much the corresponding operator accounts for the total cost.
As presented in Figure~\ref{fig:geqo-stack}, selecting a stack highlights the related operator node in red and displays its cost and percentage at the bottom.
When the selected gene is newly generated in the current generation, the \textbf{crossover visualization} in \colorbox{black}{\textcolor{white}{D4}} is activated.
An arc diagram shows how parent genes' join sequences contribute to the offspring's join sequence using color-coded edges.
Moreover, users can rerun the same query with prior optimization results and customized configuration in \colorbox{black}{\textcolor{white}{D5}} (See Section~\ref{sec:guided-geqo} for details).

These dedicated visualizations and interactive features make the complex and opaque GEQO process accessible and understandable.
Users can observe the optimizer's evolutionary behavior, understand why certain join sequences are favored, analyze cost distributions, and explore how crossover operations are executed.

\subsection{Backend}
The \systemname backend connects the frontend GUI with PostgreSQL through an Application Protocol Interface service.
As shown in Figure~\ref{fig:arc}, the query submitter transmits user queries to PostgreSQL, where they undergo parsing, rewriting, and optimization to determine the optimal execution plan.
The executor then runs the execution plan, and relevant log data are returned to the backend.
The backend parser processes this data and formats it into JSON, including all necessary fields for visualization in the user interface.

To enhance transparency into the optimization process, \systemname introduces a minimal patch to PostgreSQL that enriches its logs with detailed optimizer state data.
For the standard optimizer, the patch captures information generated during query plan enumeration, including access paths, path costs, and join sequences.
In the case of GEQO, it records each step of join sequence generation and crossover operations.
These internal states are written to a log file during query optimization.
Upon receiving query results from PostgreSQL, the backend processes the log file.
Since the log contains unstructured plain text, the parser is crucial for extracting and converting the data into a structured, readable format.
It first identifies the optimizer type and then extracts the corresponding data:
path lists, costs, and the cheapest paths for the standard optimizer; generation information, join sequences, and associated costs for GEQO.
The parsed data are formatted into JSON, combined with the query result, and passed to the frontend for visualization.

\begin{figure}[t]
    \centering
    \begin{subfigure}[t]{0.9\columnwidth}
        \centering
        \begin{lstlisting}[language=SQL]
/*+
    SeqScan(sales_returns)
    NestLoop(sales_returns catalog_page date_dim)
*/
        \end{lstlisting}
        \caption{Hint}
        \label{fig:hint}
    \end{subfigure}

    \vspace{0.5em}
    
    \begin{subfigure}[t]{0.5\columnwidth}
        \centering
        \begin{tikzpicture}
            \node[inner sep=0pt, rounded corners=5pt, clip]
                {\includegraphics[width=\linewidth]{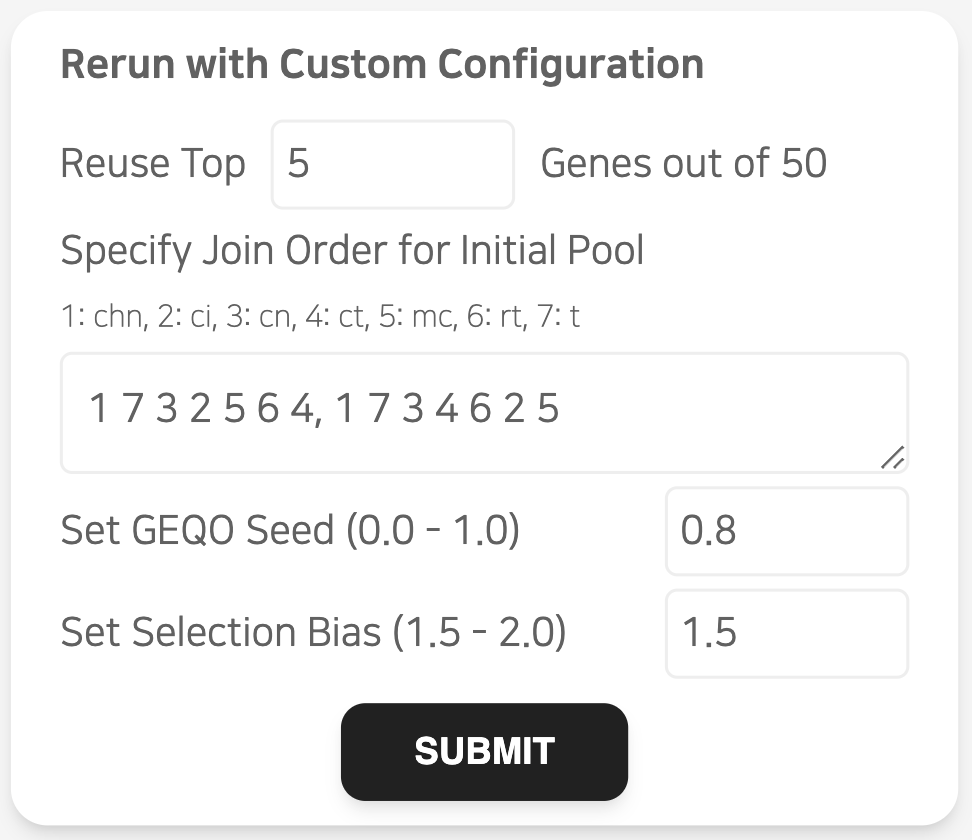}};
        \end{tikzpicture}
        \caption{Guided GEQO}
        \label{fig:ageqo}
    \end{subfigure}
    
    \caption{User-Guided Optimization}
    \label{fig:user-guided}
\end{figure}

\subsection{User-Guided Optimization}
\systemname empowers users to actively participate in the query optimization process through user-guided optimization.
By influencing the optimizer's decisions or reusing previous optimization results, users can diagnose suboptimal plans, explore alternative plans, and improve query performance in a controlled and intuitive manner.

\subsubsection{Hint}
\systemname supports the use of hints to guide the optimizer by specifying preferred physical operators or join orders.
This capability is useful in scenarios where the optimizer selects suboptimal plans or when manual tuning is required.
By providing hints, users can override the optimizer's decisions and visually analyze the impact of hints on the optimization process.
Since PostgreSQL does not natively support hints, \systemname integrates the \code{pg\_hint\_plan} extension~\cite{pg-hint} by applying a minor patch to add logging capabilities to the extension.
Hints are specified using a special comment syntax, prefixed with \code{/*+} and ending with \code{*/}, as shown in Figure~\ref{fig:hint}.

\begin{example}
    \colorbox{black}{\textcolor{white}{C}} in Figure~\ref{fig:full} demonstrates user-guided optimization using hints.
    In this example, users apply a hint to TPC-H Query 14 to enforce a nested loop join (right DAG), overriding the optimizer's choice of a hash join (left DAG).
    This change in the join method also impacts the access paths for base relations.
    In the hash join plan, the optimizer sequentially scans the \code{part} relation to build a hash table and then probes it with the \code{lineitem} relation.
    In contrast, the nested loop join plan iterates over each record in the \code{lineitem} relation and looks up matching rows in the \code{part} relation.
    The join condition is \code{l\_partkey = p\_partkey}, where \code{p\_partkey} is the primary key of the \code{part}.
    PostgreSQL automatically creates an index on primary key columns, allowing the optimizer to perform an index scan on \code{p\_partkey}.
    As a result, for each row in \code{lineitem}, matching records can be efficiently retrieved from \code{part} using the index.
    As such, users can experiment with different execution plans and observe their impact on the optimization process.
\end{example}

\begin{figure}[t]
    \centering
     
    \begin{subfigure}{0.45\columnwidth}
        \begin{tikzpicture}
            \node[inner sep=0pt, rounded corners=5pt, clip]
                {\includegraphics[width=\linewidth]{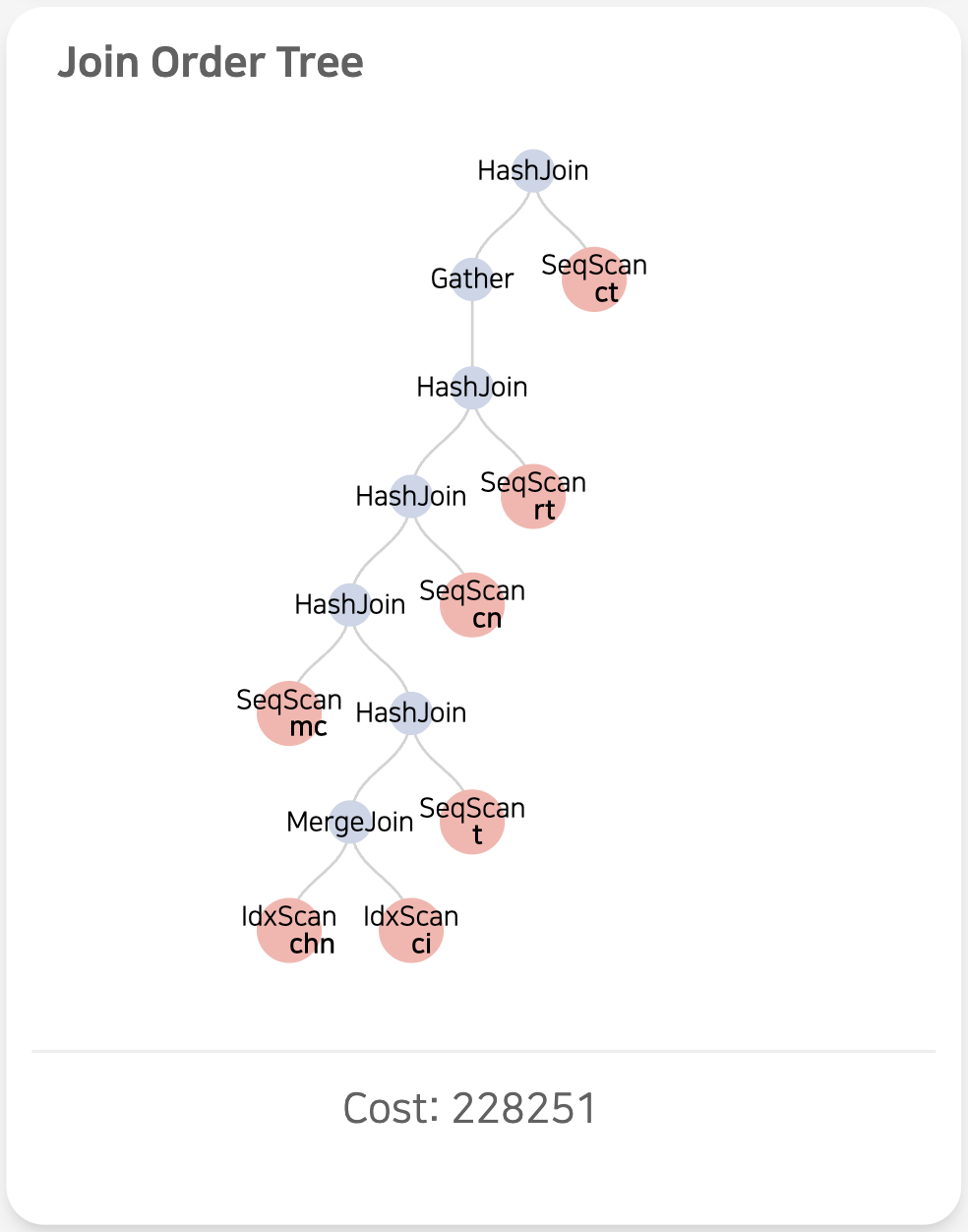}};
        \end{tikzpicture}
        \caption{Default Join Order}
        \label{fig:ageqo-before}
    \end{subfigure}
    \hspace{0.5em}
    \begin{subfigure}{0.45\columnwidth}
        \begin{tikzpicture}
            \node[inner sep=0pt, rounded corners=5pt, clip]
            {\includegraphics[width=\linewidth]{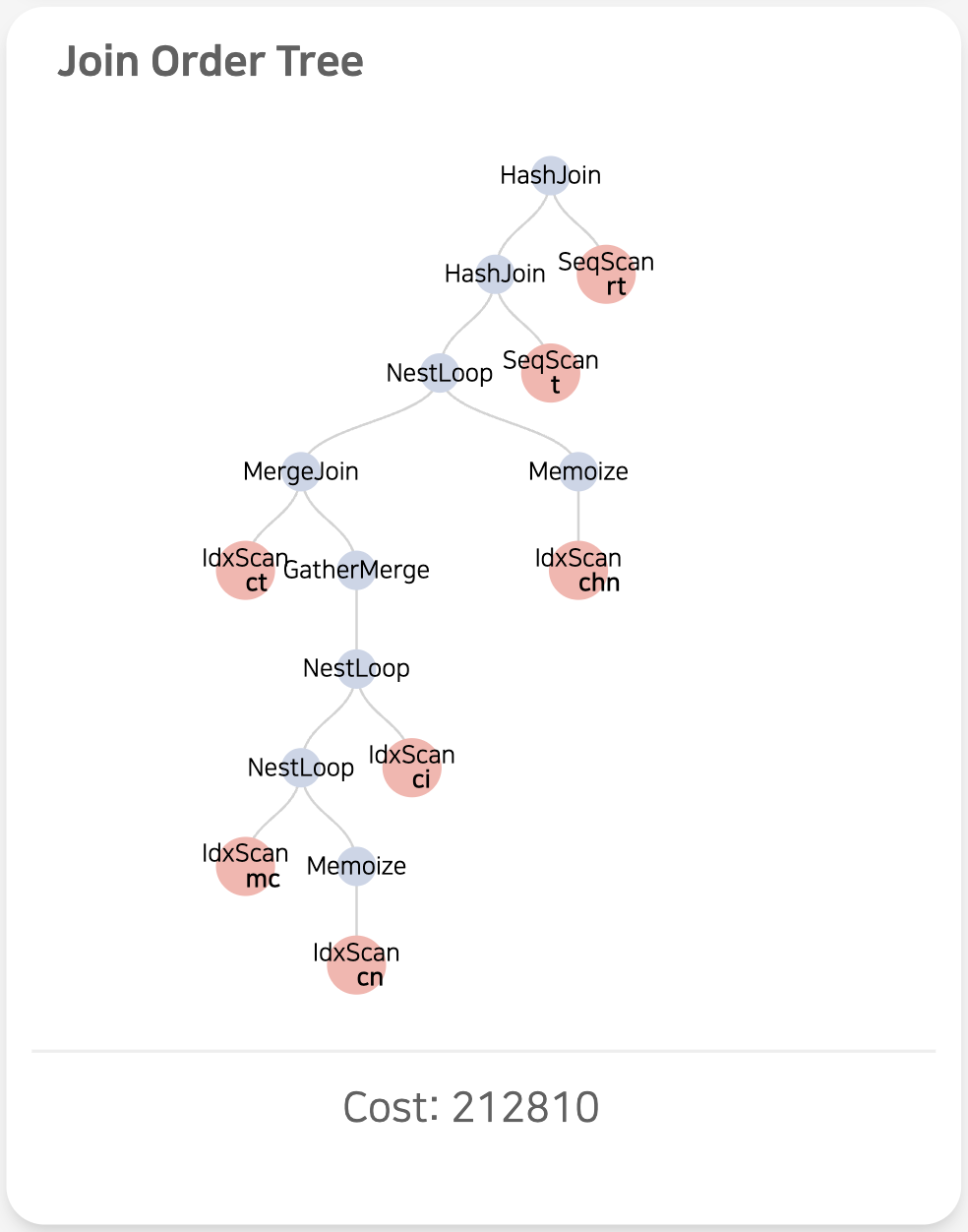}};
        \end{tikzpicture}
        \caption{User-Guided Join Order}
        \label{fig:ageqo-after}
    \end{subfigure}
    
    \caption{Guided GEQO in JOB Query 10c}
    \label{fig:demo-geqo}
\end{figure}

\subsubsection{Guided GEQO}
\label{sec:guided-geqo}
\systemname enhances GEQO by providing custom initialization of its pool.
This feature is especially useful for analytical workloads, where the same queries are often executed repeatedly.
By leveraging prior optimization results and tuning configurations, users can improve the optimality of the execution plan and explore the larger search space.

From the first two input fields in Figure~\ref{fig:ageqo}, users can customize the initialization of the pool.
They can specify the number of high-performing genes to include (\eg, the top 5 join sequences with the minimum cost from a prior run) or provide user-defined join sequences.
Above the second input field, each relation's unique ID and name are given.
Users can define join orders by entering space-delimited gene IDs and specify multiple join orders by concatenating them with commas.
This guided initialization ensures that the genetic algorithm begins with a combination of proven solutions and user-defined candidates, increasing the likelihood of discovering better execution plans while avoiding the inefficiencies of random initialization.

\systemname also provides a user-friendly interface for controlling key GEQO parameters, eliminating the need to manually add SQL commands or edit the PostgreSQL configuration file.
From the last two input fields in Figure~\ref{fig:ageqo}, users can adjust the \code{geqo\_seed}, which controls the random initialization of the pool to ensure reproducibility, and the \code{selection\_bias}, which determines the selective pressure applied during parent selection in crossover.
By tuning these parameters, users can balance exploration (\ie, maintaining diversity among genes) and exploitation (\ie, intensifying the search around high-fitness genes).

\begin{example}
    To show the effectiveness of user-guided optimization in GEQO, we use JOB Query 10c as an example.
    Figure~\ref{fig:ageqo-before} shows the query plan chosen by GEQO and its associated cost in the default configuration, with \code{geqo\_seed} set to 0 and \code{selection\_bias} set to 2.
    By rerunning the query using \systemname's user-guided optimization feature as shown in Figure~\ref{fig:ageqo}, users set a different \code{geqo\_seed} value, focus on diversity by lowering \code{selection\_bias} to 1.5, and initialize the population with the fiver lowest-cost genes from the previous run.
    As a result, GEQO discovers a more efficient execution plan, reducing the cost by 7\%, as seen in Figure~\ref{fig:ageqo-after}.
    This shows \systemnamens's ability to improve query performance through guided exploration of join sequences in GEQO.
\end{example}


\section{Demonstration}
\label{sec:demo}
In the demonstration, we will showcase the user experience with \systemname through two representative scenarios.
The demonstration highlights the effectiveness of \systemname in both educational and practical contexts, helping users understand and participate in query optimization.
In each scenario, users begin by selecting a query from the \textbf{Presets} or \textbf{History} panel in \colorbox{black}{\textcolor{white}{A}} or submitting a custom query through \colorbox{black}{\textcolor{white}{B}} and then choose the target database.
Once the query is processed, users can explore tailored optimization visualizations in the \textbf{Query Planning} tab and view the final execution plan by switching to the \textbf{\code{EXPLAIN}} tab.
In the \textbf{\code{EXPLAIN}} tab, users can zoom in and out, reposition the tree, and collapse or expand nodes by clicking on them.
This interaction is to simplify complex plans by focusing on specific parts of the execution plan for closer inspection.
Users can customize the query view layout and perform side-by-side analysis of multiple query plans using the \textbf{Queries} panel in \colorbox{black}{\textcolor{white}{A}}.
By toggling the checkboxes, users can selectively show or hide views for specific queries, and by clicking the trash icon, they can permanently remove a query view.

\subsection{Standard Optimizer}
\label{sec:demo-dp}
In scenarios where dynamic programming is applied for query optimization, \systemname constructs a DAG in \colorbox{black}{\textcolor{white}{C1}}.
The DAG presents the scan methods for base relations, join orders, and join strategies considered during query planning.
Users can zoom in and out and reposition the DAG to focus on specific portions of the planning.
By clicking the \textbf{Play} button, users can simulate the step-by-step construction of the final execution plan and observe how the optimizer selects the cheapest paths based on estimated costs.
Toggling the \textbf{Show Cost Scale} adjusts the size of operator nodes according to their estimated cost, helping users identify cost-intensive operations at a glance.
For deeper analysis, users can click on a node to examine detailed cost information in \colorbox{black}{\textcolor{white}{C2}}, including the cost factors, calculation formulas, and the values used in the calculation.
Selected nodes are highlighted with a red outline and can be selected or deselected individually.
Users can select multiple nodes to compare cost calculations across operators.
For user-guided optimization using hints, users can add directives in the \textbf{SQL editor} \colorbox{black}{\textcolor{white}{B}}.
Hints are written at the top of the query, enclosed within \code{/*+} and \code{*/}, and consist of physical operators and relations in parentheses, separated by whitespaces.
After submitting the query with hints, users can observe how the optimizer incorporates their guidance into the plan.

\subsection{Genetic Query Optimizer}
\label{sec:demo-geqo}
In scenarios where GEQO is applied, \systemname provides an interactive heatmap in \colorbox{black}{\textcolor{white}{D1}} to visualize the evolution of join sequences, represented as genes, over generations.
Users can observe how GEQO explores the search space and progressively converges toward plans with better fitness, reflected by lower estimated costs.
The \textbf{Cost Chart} \colorbox{black}{\textcolor{white}{D2}} tracks overall cost trends across generations using various metrics.
Users can hover over lines to view precise cost values at specific generations and selectively display metrics by clicking the legend.
For deeper exploration, users can hover over a gene within the heatmap to view a tooltip displaying its generation, fitness rank, and estimated cost.
Upon selection, \systemname presents a detailed view in \colorbox{black}{\textcolor{white}{D3}} that includes the join order and physical operators in a tree, along with a stacked bar chart that breaks down the estimated costs by the operator.
Users can further focus on specific generations or subsets of genes by adjusting range sliders in \colorbox{black}{\textcolor{white}{D1}}.
When the selected range is sufficiently small, \systemname visualizes the parent-child relationships produced by crossover using edges.
By selecting an offspring gene, crossover visualization \colorbox{black}{\textcolor{white}{D4}} traces the origin of each edge in its join sequence, revealing how sequences are inherited or mutated.
If the edge is inherited from both parents, it is displayed as a split color, while mutated edges are shown in gray.
For guided GEQO, users can rerun the query by specifying the top $N$ genes to initialize the pool, defining join orders if desired, and adjusting \code{geqo\_seed} and \code{selection\_bias} parameters directly in \colorbox{black}{\textcolor{white}{D5}}.

\bibliographystyle{ACM-Reference-Format}
\bibliography{references}

\end{document}